\documentclass[10pt]{amsart}
\usepackage{geometry}                
\usepackage{amssymb}

\usepackage{graphicx}
\usepackage{epstopdf}
\usepackage{url}
\usepackage{dcolumn}
\usepackage{bm}
\usepackage{color}
\usepackage{mhchem}
\usepackage[foot]{amsaddr}

\renewcommand{\vec}[1]{{\mathbf{#1}}}
\newcommand{\tens}[1]{{\mathbf{\overset{\leftrightarrow}{#1}}}}
\newcommand{\angy}[1]{{\mathbf{\widetilde{#1}}}}

\newcommand{\NA}{\mathrm{NA}}

\newcommand{\refl}{\mathcal{R}}
\newcommand{\trans}{\mathcal{T}}
\newcommand{\abs}{\mathcal{A}}

\newcommand{\units}[1]{{\,\mathrm{#1}}}

\newcommand{\fix}[1]{{#1}}

\title{Rapid mapping of digital integrated circuit logic gates via multi-spectral backside imaging}
\author{Ronen Adato$^{1,4,*}$, Aydan Uyar$^{1,4}$, Mahmoud Zangeneh$^{1,4}$, Boyou Zhou$^{1,4}$, Ajay Joshi$^{1,4}$, Bennett Goldberg$^{1,2,3,4}$ and M Selim \"{U}nl\"{u}$^{1,3,4}$}

\address{$^1$Department of Electrical and Computer Engineering, Boston University, Boston, MA 02215\\
$^2$Physics Department, Boston University, Boston MA 02215 \\
$^3$Department of Biomedical Engineering, Boston University, Boston MA 02215\\
$^4$Photonics Center, Boston University, Boston MA 02215}

\begin{document}

\date{\today}

\begin{abstract}
Modern semiconductor integrated circuits are increasingly fabricated at untrusted third party foundries. There now exist myriad security threats of malicious tampering at the hardware level and hence a clear and pressing need for new tools that enable rapid, robust and low-cost validation of circuit layouts. Optical backside imaging offers an attractive platform, but its limited resolution and throughput cannot cope with the nanoscale sizes of modern circuitry and the need to image over a large area. We propose and demonstrate a multi-spectral imaging approach to overcome these obstacles by identifying key circuit elements on the basis of their spectral response. This obviates the need to directly image the nanoscale components that define them, thereby relaxing resolution and spatial sampling requirements by 1 and 2 - 4 orders of magnitude respectively. Our results directly address critical security needs in the integrated circuit supply chain and highlight the potential of spectroscopic techniques to address fundamental resolution obstacles caused by the need to image ever shrinking feature sizes in semiconductor integrated circuits.
\end{abstract}

\maketitle

\section{Introduction}\label{introduction}

Semiconductor integrated circuits (ICs) are pervasive and essential components in virtually all modern devices, from personal computers, to medical equipment, to varied military systems and technologies. 
Their functionality is defined by a massive number ($\sim 10^6 - 10^9$ currently) of interconnected logic gates that correspond physically to various nanoscale doped regions, polysilicon and metal (usually copper and tungsten) structures.
Modern ICs are thus extraordinarily complex physical structures that are produced by equally complex processes.
Validating and testing chips is becoming both increasingly important and challenging \cite{ITRS,Ma2012,Vigil2014}.
New techniques are essential from both a failure analysis and, in recent years, a security standpoint \cite{DSB2005,Chakraborty2009,Bhunia2014,Rostami2014}.

Security issues have become significant as a result of the increasingly complex, fragmented and global multi-stage process by which ICs are produced.
This trend has opened the door for numerous security threats including piracy, counterfeiting \cite{Ahmood2015} and malicious tampering \cite{DSB2005,Chakraborty2009,Bhunia2014,Rostami2014}.
Malicious tampering is manifest by the insertion of a few rogue gates into an IC in order to, for example, subvert firewalls, leak sensitive information or compromise device functionality \cite{DSB2005,Chakraborty2009,King2008,Bhunia2014,Rostami2014}. 
In analogy with the software threats, these are termed Hardware Trojans \cite{Chakraborty2009,Bhunia2014}.
Unlike software vulnerabilities, Hardware Trojans require highly specialized equipment and expertise to detect and cannot simply be patched in the field.
Efforts by the U.S. Defense Department to control the full process for fabricating its chips in the Trusted Foundry program \cite{DSB2005} and the large body of research on possible modifications to the chip design and fabrication \cite{Rajendran2012,EllMassad2015,Imeson2013,Rajendran2013,Valamehr2013} flow highlight the significance of this problem as well as the lack of appropriate testing solutions.

Current electronic testing methods fall short in testing for Hardware Trojans. Digital tests cannot exhaustively sample the massive state space which scales nonlinearly with IC complexity, and analog measurements are highly sensitive to minor fabrication variations \cite{King2008,Rostami2014,Bhunia2014}. Both can be readily designed around by an adversary \cite{King2008,Rostami2014,Bhunia2014}.
There is therefore a clear and pressing need for tools that enable direct, rapid and low cost detection of IC tampering.

Any change in the functionality of an IC must map to a change in the physical structures that define its logic.
A direct image of the layout therefore offers the most direct and exhaustive route to detect tampering.
Measuring the physical structures as opposed to the electrical signals produced by an IC is fundamentally advantageous.
Electronic measurements must consider and be designed to be robust over a wide range of complex activation and action characteristics of Hardware Trojans \cite{Bhunia2014}, in contrast to direct IC layout images that do not.
Spatial mapping also has more favorable scaling rules since the measurement space varies only linearly with transistor or gate count.
Traditional imaging methods, however, are fundamentally limited in their ability to handle the enormous range in length scales that are present in modern ICs.
The physical features that define the transistors and their associated interconnects have dimensions below 100 nm, yet billions of them together cover an area on the order of $\mathrm{cm^2}$.
From the standpoint of the optical resolution, these dimensions are at the limits of what can be achieved with solid immersion lenses (SILs) at the near-infrared (IR) wavelengths ($1 - 3 \,\mu\mathrm{m}$) that are required to optically access IC circuitry through the silicon (Si) substrate \cite{Ippolito2001,Serrels2008b,Koklu2009,Vigil2014,Agarwal2015}. 
The extremely high numerical apertures ($\NA$) required, however, sacrifice field of view and thus present significant challenges to imaging over large areas.
A similar bottleneck exists in terms of the number of spatial samples that would be required to image a full IC at a discretization commensurate with the smallest feature sizes, which are $\sim10 - 100 \units{nm}$.
To simply record this many image points (pixels) with a conventional laser scanning confocal microscope would require at least several hours to tens of days.

In this work we propose and theoretically demonstrate the use of a multi-spectral imaging approach that takes advantage of the inherent modularity of IC design to overcome these fundamental obstacles.
The basis for our method is that physical structure of a digital IC maps to a well defined series of interconnected logic gates.
Each gate type is physically implemented on the chip through a series of distinct metal wire and transistor geometries that define the appropriate electronic behavior.
The physical layout of a digital IC thus takes the form of a tiling of a few different types of standard cells that, significantly, have linear dimensions on the order of microns.

Imaging at the resolution commensurate with these micron scale gates, as opposed to the nanoscale wires and transistors that define them allows for dramatic reductions in spatial resolution and sampling requirements.
Instead of identifying these gates on the basis of images of their detailed sub-structure, we show here that a small number ($\sim 5$) of measurements at different wavelengths and polarizations allow one to distinguish between gates with near 90\% accuracy via a simple pattern recognition approach.
This enables the possibility of mapping the type and location of every logical gate in an IC in a few minutes using a very modest 0.8 NA imaging system.
The method we propose here thus directly addresses critical security issues in the IC pipeline and is highly scalable to future ITRS projections.
We demonstrate our method with numerical simulations of a backside optical imaging system such as those used for IC failure analysis and a 45 nm technology node open source library \cite{Nangate45} as a test case.

\begin{figure}[htb]
  \centering
  \includegraphics[width=0.5\textwidth]{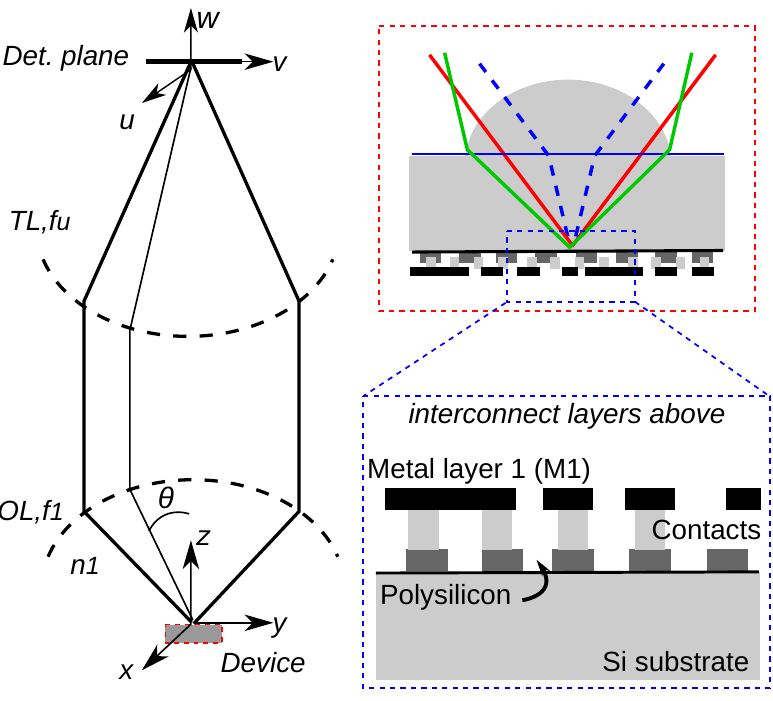}
  \caption{Integrated circuit backside imaging.
          A diagram of a generalized image forming system and coordinates along with detail of backside imaging geometries are shown. The Gaussian reference spheres for the objective ($OL$) and tube ($TL$) lenses are shown as the black dashed curves in the diagram. The coordinate systems are as used in all calculations. The upper inset (red-dashed) illustrates the ray paths for a conventional air objective (blue dashed lines), central (red) and aplanatic (green) solid-immersion lens objectives.
          The lower inset (blue-dashed) shows detail of the active, polysilicon, contact and local metal (M1) layers in a typical integrated circuit.
          Layers above the substrate are embedded in an insulating dielectric, taken to be silicon-dioxide ($\mathrm{SiO_2}$) here}.
  \label{fig:image-geometry}
\end{figure}


\section{Backside Imaging of Integrated Circuits}

Integrated circuits consist of a layered stack built up over a Si substrate.
The bottom most layers define the active region containing the transistors and are comprised of doped regions in the Si and polysilicon.
The first metal layer (M1) typically is used to connect combinations of these transistors so as to form the standard CMOS logic gates.
Imaging these most basic structures would provide an ideal means with which to directly verify the faithful fabrication of a submitted chip layout at an untrusted foundry.
However, achieving the required resolution and throughput is extremely challenging.

In a standard scanning confocal microscope the object is illuminated by a focused spot to produce an induced polarization, or current distribution $\vec{j}(x, y, z)$, in the object. 
On the collection side the radiation scattered in the far-field due to this induced current is imaged onto the detector plane according to, \cite{Novotny2006}

\begin{align}\label{eq:image_fields_psf}
  \vec{E}_D(u, v) &= \iiint_V \tens{G}(u, v; \vec{r}) \vec{j}(\vec{r}) \,d^3\vec{r}
\end{align}

\noindent where $\vec{r} = [x, y, z]$ is the object space position, $\tens{G}$ is the Green's tensor \cite{Novotny2006} for the imaging system, $\vec{E}_D$ is the electric field at the detector plane ($w = 0$).

A pixel value at position $(a,b)$ is recorded by integrating the intensity at the detector plane over an aperture, $\mathcal{W}$.
This is general a pinhole similar in size to the point spread function at the detector plane.
The resultant signal, $S(a, b)$, can be written as,

\begin{align} \label{eq:image_response_pixels}
  S(a, b) &= \alpha\iint_{\mathcal{W}} \left\vert \vec{E}_D(u, v) \right\vert^2
                                          \, du \, dv
\end{align}

\noindent where $\alpha$ is a scaling constant.
The sample is raster scanned, usually by means of a scanning galvo and relay system, such that every $(\Delta_x, \Delta_y)$ a new pixel is recorded to build up the image.

Optical resolution limits are dictated by the fundamental physics described in equation~\ref{eq:image_fields_psf}.
As an estimate, consider the paraxial limit where the dominant component of $\tens{G}$ takes the form of an Airy disc whose size is approximately $0.61\lambda/\NA$, where $\NA$ is the numerical aperture of the collection objective.
The blurring of the object that results leads to the well known Abbe diffraction limit.
Imaging the IC structures through the Si substrate requires using near-IR ($\lambda = 1 - 3 \,\mu\mathrm{m}$) light.
This therefore severely limits resolution given that conventional air objectives have $\NA$s below $1$.
For a reasonable, $\NA = 0.8$ system, Abbe's approximation indicates a resolution of $\sim 750 \units{nm}$ at $\lambda = 1 \,\mu\mathrm{m}$, significantly larger than the sizes of the relevant IC wires.

Solid immersion lenses (SILs) address these issues by placing a Si hemisphere in contact with the IC substrate backside.
Central or aplanatic configurations preserve or increase, respectively, the angles of incident rays as they pass into the Si substrate enabling $\NA$s as high as $n_{Si} = 3.5$ theoretically.
Recent research aimed to address IC failure analysis needs has pushed the resolution limits of traditional solid immersion lens subsurface microscopy to $\sim 100 - 150 \units{nm}$ \cite{Koklu2009,Serrels2008b,Vigil2014,Agarwal2015}.

Nevertheless, typical (minimum) line widths and separations in the M1 layer are 65 nm for the 45 nm process we use here to validate our results.
These features will therefore be just at or below the impressive resolution limits of SIL systems.
At leading edge, 22 nm and below, technology nodes \cite{ITRS} the mismatch will be even worse.

Throughput limitations present an additional significant challenge.
In contrast with optical resolution limits, they do not the result from a fundamental physical process such as diffraction.
They are due to the need to sample at rates commensurate with the smallest, nanoscale, structures to be imaged, yet map over a centimeter scale area.
As a simple quantitative estimate, a reasonable sampling rate of $\Delta = 10 - 100 \units{nm}$, implies that $10^{12} - 10^{10}$ samples would be needed to cover a $\mathrm{cm^2}$ area.
At the MHz acquisition rates that are characteristic of conventional galvo scanning confocal systems several hours to tens of days would be required to scan a chip size area.

\section{Gate-level Mapping of Integrated Circuits}

\begin{figure}[h!tb]
  \centering
  \includegraphics[width=0.5\textwidth]{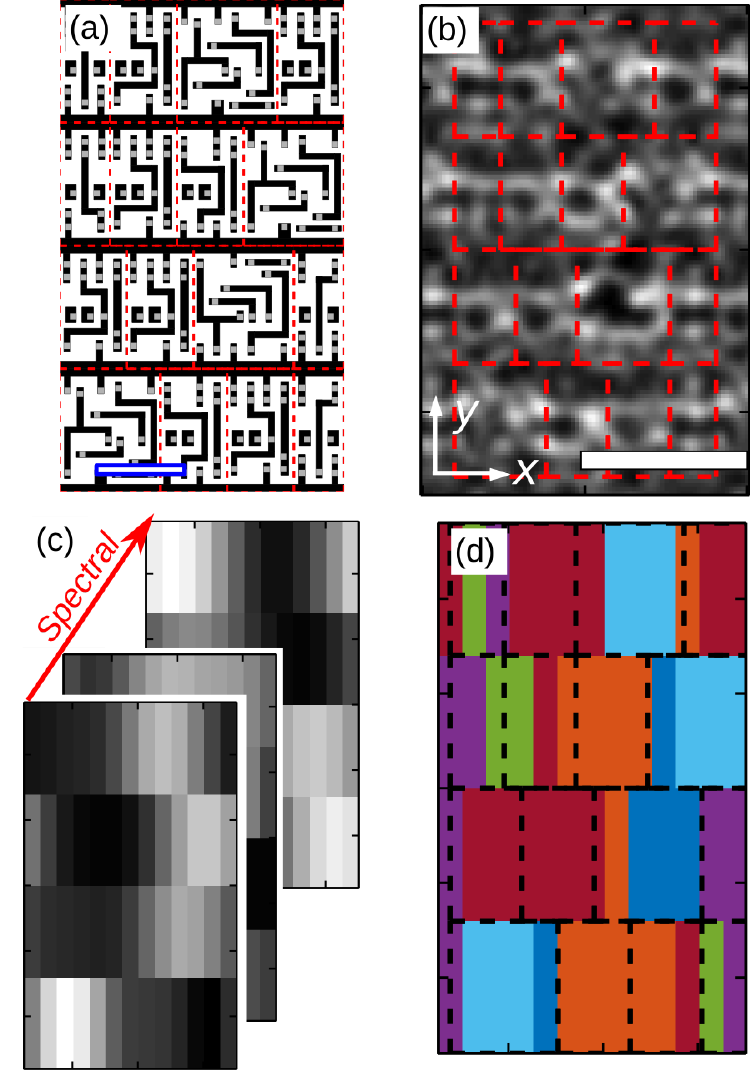}
  \caption{Multi-spectral backside imaging for rapid gate-level mapping  of an integrated circuit.
          (a) Layout of a 4 x 4 tiling of integrated circuit gates (black lines and gray squares correspond to M1 lines and contacts respectively). Red dashed lines indicate gate boundaries. 
          (b) Simulated high NA (3.4) image of the layout at a wavelength of $\lambda = 1060 \units{nm}$. The circuit is illuminated with light polarized in the $y$-direction. 
          (c) Multi-spectral image cube of the circuit. The images are collected at low resolution (NA = 0.8 and sampled at a low rate in the spatial dimension ($\Delta = 250$ and 1400 nm in the $x$ and $y$-directions). Each panel corresponds to an image collected at a different wavelength-polarization combination. 
          (d) Gate map generated from the data cube in (c). Pixels are colored to indicate one of 6 possible gates.
          The scale bars in (a) and (b) are 1 and $2\, \mu\mathrm{m}$ respectively.}
  \label{fig:image-concept}
\end{figure}

The numerical simulations in Figure~\ref{fig:image-concept} highlight the challenges of traditional imaging methods and illustrate our proposed solution.
Full details of our computational method are given in the Methods section (Appendix A1 - A3).
In brief, we use a finite-difference time-domain (FDTD) solver \cite{Lumerical} to simulate the scattering of an incident beam by the nanoscale IC circuit components.
The far-fields are be obtained directly from these simulations via the near-to-far-field transform and then propagated through our imaging system using the angular spectrum approach and Debye approximation \cite{Richards1959,Davis2010,Davis2010b}.

We neglect refraction caused by light exiting the Si substrate for simplicity and note that this is easily avoided experimentally by commonly employed and commercially available central SIL microscopes. 
While all calculations are therefore strictly representative of this geometry we expect that the general concepts we develop here can be utilized with a variety of different optical configurations including without a SIL.

Figure ~\ref{fig:image-concept}a shows a the M1 and contacts (black and grey regions respectively) associated with a series of logic gates taken from the 45 nm Nangate library~\cite{Nangate45}.
Their image produced using a 3.4 NA system \cite{NA_note} at a wavelength of $\lambda = 1060 \units{nm}$ and 10 nm sampling rate is shown in Figure~\ref{fig:image-concept}b.
Even at such high resolution and sampling rate the detailed geometry of the metal wires cannot be resolved.
With neither adequate resolution, nor sufficient throughput, traditional imaging methods cannot offer a viable solution to IC assurance and security.

While the individual metal wires shown in Figure~\ref{fig:image-concept}a are several 10s of nanometers in scale, the digital logic gates that they ultimately define are over an order of magnitude larger.
The boundaries of these gates are outlined in red in the figure.
For the particular 45 nm process we examine here their linear dimensions are on the order of $1\units{um}$.
Treating the micron scale gates as opposed to the wires and transistors that define them as the smallest spatial unit to image therefore offers a route to dramatically relaxing resolution and sampling rate requirements.
Optical resolution on the order of $1 \,\mu \mathrm{m}$ can easily be achieved even without a SIL.
Sampling at this rate such that approximately one gate maps to one pixel reduces the required pixel count by 2 - 4 orders of magnitude compared to the previous estimates.
This implies the potential to image a full chip ($\mathrm{cm^2}$ area) in 2-3 minutes considering again MHz acquisition rates.

At such low resolution and pixel count the image itself cannot be used to distinguish between the different gates.
We therefore sought to instead identify the gates on the basis of their response to light incident at different wavelengths and polarizations (referred to as spectral response hereafter for brevity).
Spectral scattering and absorption has been used to identify and distinguish between particulates \cite{Lindfors2004,Sonnichsen2005}, tissue class in histopathology applications \cite{Fernandez2005,Baker2014} as well as characterize the sub-wavelength resonant modes present in designer metamaterials \cite{Fan2010}.
We apply similar principles here.
We intentionally illuminate and image our sample with a low NA objective to match the spot size with the gate dimension (0.8$\NA$ is used here and in all the following).
As a result, the individual wires that comprise each gates are excited uniformly and their collective response measured.
This allows us to roughly think of each gate as an effective material that can be identified on the basis of their overall spectral response.
As shown in the results in Figures 2c,d, a series of low-resolution, coarsely sampled images taken at a few (5) different spectral features can be used instead of the high resolution, finely sampled one.
Applying established pattern recognition techniques \cite{Duda2012,Fernandez2005} allows us to associate each spatial pixel in the image stack with a different logic gate to form the final map in Figure~\ref{fig:image-concept}d.
In comparison with the purely spatial sampling approach, these 5 spectral measurements are dramatically more efficient than the $\sim 10^2-10^4$ per gate (i.e. $1 \, \mu \mathrm{m^2}$ area) that would be needed to with $\Delta = 10 - 100 \units{nm}$.

Our approach represents a new and unique application of spectroscopic imaging.
In addition to the first application of spectral imaging to IC backside imaging, the overarching goal and motivation is distinct from traditional applications such as those in biotechnology and remote sensing.
Rather than seeking to obtain new mechanisms for contrast based on sample specific spectral bands, our aim is to overcome the need for impossibly high resolution and spatial sampling rates by more efficiently probing structure in the spectral domain.

\begin{figure}[htb]
  \centering
  \includegraphics[width = 0.6\textwidth]{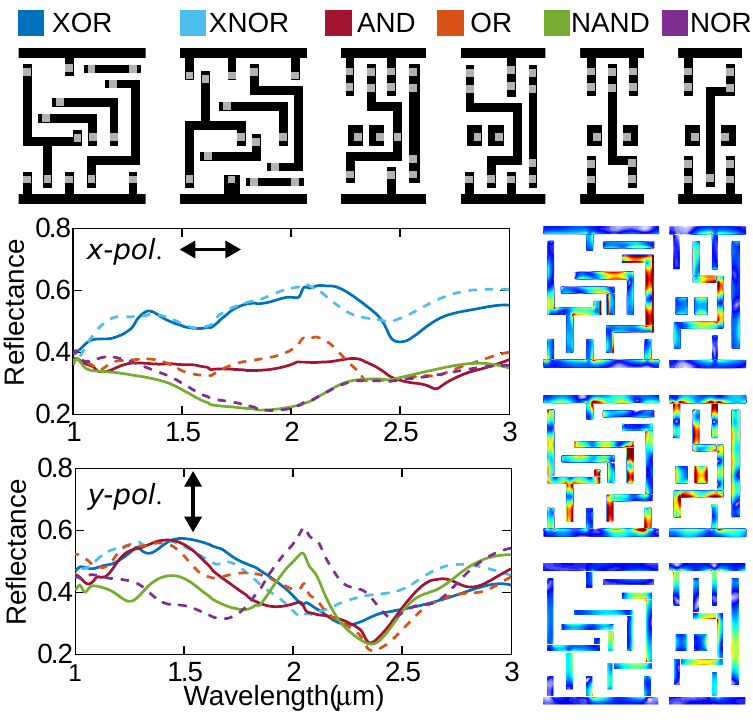}
  \caption{Spectroscopy of integrated circuit `metamaterials'.
          The top row indicates the legend and shows schematics of the metal 1 (black lines) and contact (gray squares) layers of the six gates.
          The plots show simulated spectral response for the 6 gates under x- and -y polarized illumination. 
          Images of the induced current along the wires in the XOR and AND gates at various wavelength ($\lambda$) and polarizations ($q$) are shown on the right. The $(\lambda, q)$ values are (1500 nm, $x$), (1100 nm, $y$) and (2600 nm, $y$) from top to bottom.
          }
  \label{fig:gate_spectra}
\end{figure}

\subsection{Spectral Fingerprints of Logic Gate Geometries}
In the backside imaging geometry we consider, one measures the light reflected by the metal wires and contacts that define the IC gates as well as the interface between the Si substrate and insulating oxide (SiO2) in which the IC structures are embedded.
The FDTD simulations in Figure~\ref{fig:gate_spectra} illustrate the link between this particular response - the spectral reflectance - and the substructure of the different IC gates.

The results show the net reflectance of a periodic tiling each gate (see Appendix A2) under linearly polarized plane-wave illumination.
Each of the six gates in our test set is defined by a distinct series of M1 wires.
When illuminated, these \emph{wavelength scale} wires support current oscillations that are highly dependent on the wire geometry and the illumination wavelength and polarization.
From equation~\ref{eq:image_fields_psf}, the detected signals will therefore be highly \fix{distinct} for different gates and vary strongly as a function of wavelength and polarization of the incident field.
This is evident in the reflectance spectra, and series of near-field current distributions shown for the XOR and AND gates.
An important feature that is obvious from the combination of near and far-field data shown in the figure is that although the induced current distributions and far-field spectral responses are unique fingerprints of the different gates, there is no simple one-to-one mapping between specific modes and far-field features.
Instead the current modes couple to each other in a complex fashion in the near-field and produce a far-field response that is the result of the coherent interference between their scattered radiation as well as the signal reflected from the Si-$\mathrm{SiO_2}$ interface.
This motivates the statistical classification approach we will apply to distinguish between the different gates.

A second observation is that the majority of the current distributions excited appear to correspond to higher order modes.
This is important as it implies that we can expect geometry dependent spectral features to exist over an extremely broad wavelength range and to persist in the near-IR for essentially any IC gate design even as dimensions are reduced at advanced technology nodes.
We can intuitively validate this by considering a simple estimate approximating the wires as Fabry-Perot resonators for current modes in accordance with common models for dipole antennas.
The lowest order (longest wavelength) mode is at $\lambda \approx 2n_{eff}L$ ($n_{eff}$ is the effective index and $L$ the wire length) \cite{Novotny2007}.
The length scales of the M1 wires in the gates in Figure~\ref{fig:gate_spectra} range from $\sim 100 - 1500 \units{nm}$.
This indicates that a spectral response that is a strong function of the gate geometry can be observed at wavelengths between $0.4 - 9 \,\mu\mathrm{m}$ assuming $n_{eff} \approx 3$ as a result of proximity to the Si substrate.
This extremely broad range that extends far into the infrared indicates ample room for reduction in wire length before near-IR resonant structures are no longer naturally present.

\subsection{Gate Response and Variability in a Scanning Microscope}

These results indicate the significant potential for the spectral reflectance of each gate to act as fingerprint that can be used to identify them.
In practice our imaging system records a subset of the signals presented in Figure~\ref{fig:gate_spectra}.
Both the limited collection cone of the objective ($\pm13\deg$ for $0.8\NA$ in a Si background) and the pinhole at the detector plane will reject a portion of light scattered by the IC structures.
The imaging geometry also implies the need to consider the relation of each image pixel to a spatial position of the sample as it is raster scanned and the tiled logic gates that we aimed to identify.

In general, following equation~\ref{eq:image_response_pixels}, the signal at a given pixel will have the form (see Appendix A1 for derivation),

\begin{equation}\label{eq:refl_xy_main}
  S(a_i, b_j; \lambda, q)/S_0(\lambda, q) = \refl(x_i, y_j; \lambda, q) + \eta_{det}(\lambda, q)
\end{equation}

\noindent where $\lambda$ and $q$ index the wavelength and polarization state of the incident beam such that $(\lambda, q)$ defines the specific spectral feature that is measured.
In equation~\ref{eq:refl_xy_main} we have assumed that the signal has been normalized by a background, $S_0$, that removes wavelength and polarization dependent variations that are inherent to any realistic experimental system.
These include detector responsivity, source power spectral dependence and absorption in the Si substrate.
This isolates the portion of the reflectance that depends solely on the spatially varying IC structure that is of interest.
This reflectance, denoted $\refl(x_i, y_i; \lambda, q)$, is assumed to account for the full effects of the imaging system as indicated above.
It varies with the sample geometry at position $(x_i, y_i)$ and directly measured via the normalized pixel response of equation~\ref{eq:image_response_pixels}.

To link this practical measurement to the spectral reflectance of our gate objects we associate pixels $(a_c, b_c)$ that map to sample positions $(x_c, y_c)$ that coincide with the center coordinates of a specific gate $c$ with that gate.
We can therefore write,

\begin{equation}\label{eq:refl_to_ps}
  S(a_c, b_c; \lambda, q)/S_0(\lambda, q) = \refl(c; \lambda, q) + \eta_{\delta}(c; \lambda, q) + \eta_{det}(\lambda, q)
\end{equation}

\noindent where we have written $\refl(c; \lambda, q) = \refl(x_c, y_c; \lambda, q)$ for short.

Equation~\ref{eq:refl_to_ps} thus describes the measurement of the spectral reflectance of a given gate class, $\refl(c; \lambda, q)$.
The measurement depends not only the specific gate type centered under the microscope and the feature measured through $\refl(c; \lambda, q)$, but also will inevitably include fluctuations caused by two noise sources, $\eta_{\delta}$ and $\eta_{det}$.
System noise, $\eta_{det}$, is independent of the actual sample position or underlying IC structure and is manifest as detector noise.
The other noise term, $\eta_{\delta}$, originates from the fact that our measurements of a given gate will likely also contain some signal from scattering off of neighboring gates.

Because of the presence of noise, $\eta = \eta_{det} + \eta_{\delta}$, the relationship between a measurement, $S(a_c, b_c; \lambda, q)/S_0(\lambda, q)$, and the spectral reflectance of the underlying gate class, $\refl(c; \lambda, q)$, will not be deterministic.
A key consideration is therefore to determine an ideal small set of features $(\lambda_i, q_i)$ to measure that will offer maximal accuracy in distinguishing between the different gate classes.
This depends directly on the elements of the optical system that govern $\refl$, $\eta$ and the range of spectral features that can be accessed.

In order to address these issues formally we utilize Bayes' theorem, which is the basis for an intuitive and established classification method \cite{Fernandez2005,Duda2012}.
We consider a spectral reflectance measurement $M_i = S(c_0; \lambda_i, q_i)/S_0$ that we aim to use to identify the underlying class, $c_0$, via the dependence on $\refl(c_0; \lambda_i, q_i)$.
The probability that our measurement $M_i$ implies an underlying class, $c_j$ is given via Bayes' theorem, $P(c_j \vert M_i) = P(M_i \vert c_j)P(c_j)/P(M_i)$.
The distribution of measurement values, $M_i$, for each gate dictates the probability $P(M_i \vert c_j)$ and $P(c_j)$ scales for the prior probability of each class.
The denominator, $P(M_i)$, amounts to a normalization constant.
For multiple measurements, $\vec{M} = \left[M_1, M_2, \ldots, M_N \right]$,

\begin{align}\label{eq:bayes}
  P(c \vert \vec{M}) &= P(\vec{M} \vert c)P(c)/P(\vec{M}) \\
    &\approx P(M_1 \vert c)P(M_2 \vert c) \ldots 
        P(M_N \vert c)P(c)/P(\vec{M}) \nonumber
\end{align}

\noindent where the approximation made in the second line assumes statistical independence of the measurement features. 
This assumption is not strictly accurate, but it significantly simplifies the process of determining probability distributions and calculating the probabilities and is commonly applied \cite{Fernandez2005,Duda2012}.
If the probability distributions that determine the right hand side of equation~\ref{eq:bayes} can be characterized, each pixel can be classified as the specific gate type, $c$, that corresponds to the maximum $P(c \vert \vec{M})$.
This amounts to dividing the $N$ dimensional measurement space into regions $R_j$ such that $\vec{M} \in R_j \rightarrow c = c_j$.
Mis-classification errors therefore intuitively result from overlapping probability distributions.

\begin{figure}[ht!b]
  \centering
  \includegraphics[width=0.75\textwidth]{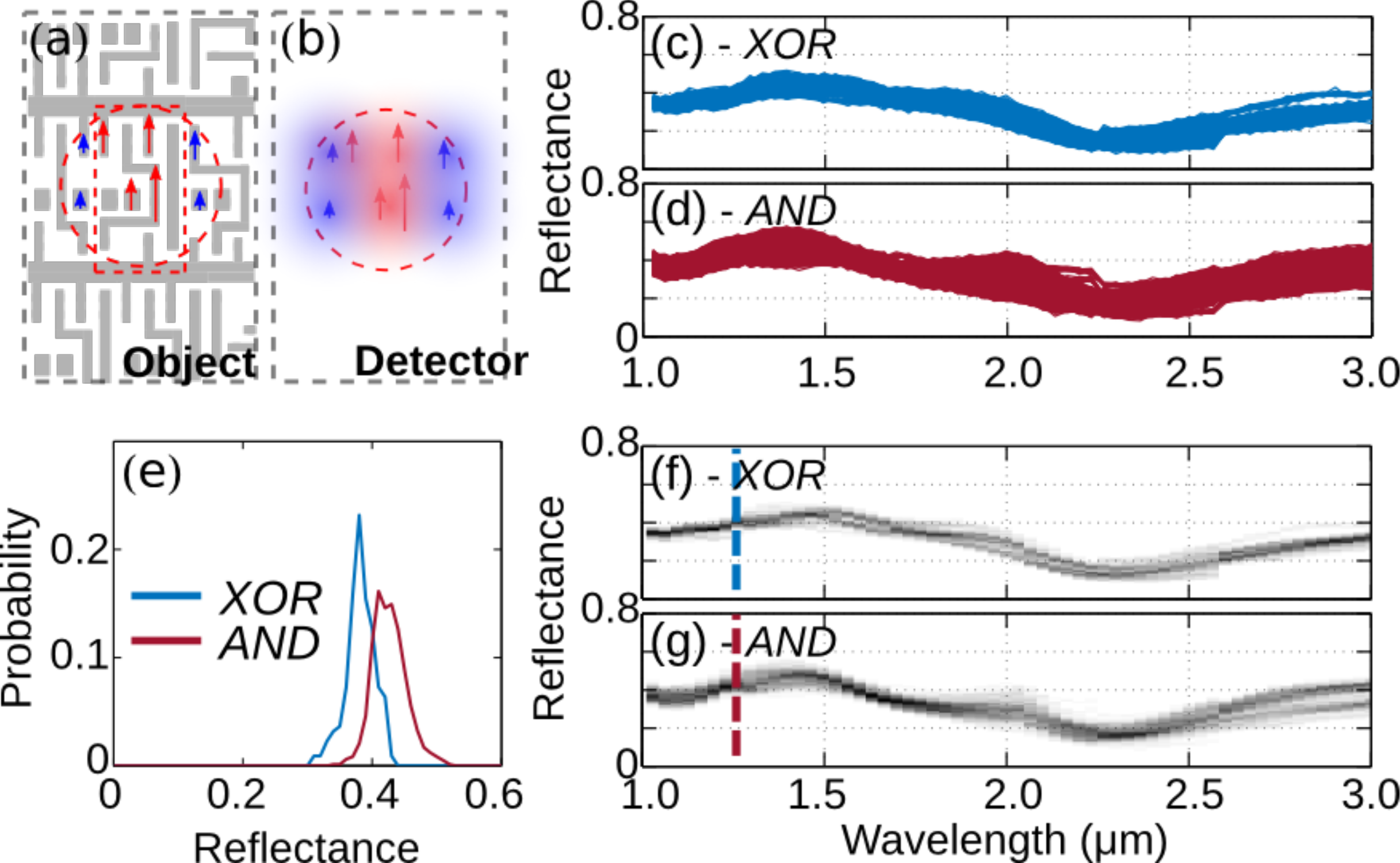}
  \caption{Sources of variation in integrated circuit gates' spectral fingerprints 
          (a,b) Illustration of cross-talk sources in low-resolution gate spectral images. 
          (a) Schematic of the excitation of currents in the metal lines by a \textasciitilde{}gate-size spot. The incident beam induces currents along the wires inside the target gate (red arrows) as well as in its neighbors (blue). 
          (b) Illustration of the signal generated via these induced currents at the detector plane. The spatially selective element, e.g.~pinhole or pixel, defines a subset of these to integrate over (red dashed circle). 
          (c,d) Variation in the spectral fingerprints of XOR and AND gates due to adjacent gates for a 0.8 NA imaging system, under \emph{y} polarized illumination.
          (e) Probability distributions for the XOR and AND gate responses at $\lambda = 1227 \units{nm}$ and $y$ polarization. 
          (f,g) Probability distributions for the two gates at all wavelengths, displayed as image plots with darker regions indicating higher probabilities. 
          \fix{The dashed blue and red vertical lines indicate the slice corresponding to panel (e).}
          } 
  \label{fig:gate_histograms}
\end{figure}

Our ability to accurately distinguish between and identify the various gates in our set therefore depend heavily on our imaging system. 
Its characteristics determine the form of $\refl$ and $\eta$ and therefore probability distributions for $P(M_i \vert c)$.
These depend on fundamental physical constraints dictated by, for example, optical resolution as well as practical considerations that govern the availability and noise characteristics of sources and detectors.

\section{Influence and sources of noise}
For the relevant near-IR spectral region on which we focus, neither detector nor source availability should place significant constraints on the spectral features that could be used as classification measurements.
Standard InGaAs detectors operate between $1 - 2.6 \,\mu\mathrm{m}$ and have response times appropriate for fast imaging.
Numerous sources exist that are suitable for suppling a moderate number of different input wavelengths.
The primary factors governing the performance of our approach therefore are sources of error.

To assess the influence of system noise and the fundamental limits on our method, we considered (i) detector noise and; (ii) cross-talk between adjacent gates.
Detector noise will always be present and sets a lower bound on measurement variation for a given measurement speed.
Cross-talk between adjacent gates is directly related to both the size of the point-spread function described in equation~\ref{eq:image_fields_psf} and the integration area in equation~\ref{eq:image_response_pixels}.
It is fundamental to our approach since it is a function of the resolution and throughput trade-offs that we aim to address.
In principle one can engineer around other sources of error such as experimental error due to sample alignment.

We approximated detector noise as resulting in Gaussian uncertainty distribution for a given measurement.
The standard deviation (in reflectance scale), $\sigma$, can be determined from the usual definitions for detector signal-to-noise ratio ($S/N$) via $\sigma = (S/N)^{-1} = P_{NEP}\sqrt{\Delta f}/P_S$, where $P_S$ is the power incident on the detector, $P_{NEP}$ is the noise equivalent power with respect to a 1 Hz bandwidth and $\Delta f$ is the measurement bandwidth \cite{Boyd1983}.
Common InGaAs detectors have $P_{NEP}\sim 10^{-12} \units{W/\sqrt{Hz}}$ such that on the order of $P_S \sim 0.1 - 1\,\mu\mathrm{W}$ signal power are sufficient for $\sigma < 0.01$ at MHz acquisition rates ($\Delta f = 10^6 \units{Hz}$), with the exact value depending on the wavelength, \fix{responsivity} curve of the detector \fix{and thickness of the Si substrate (Appendix~\ref{app:snr:pwr-budget} presents detailed calculations)}.
The separation between the spectral reflectance curves in Figure~\ref{fig:gate_spectra}b,c, is generally better than $0.05$ such that noise on the order of $0.01$ or below enables high classification accuracy.
As a quantitative estimate we calculated the total error rate analytically via an overlap integral assuming Gaussian $p(M \vert c)$ centered at the reflectance values from Figure~\ref{fig:gate_spectra}b and equal prior probabilities for each class.
Classification accuracies above $99\%$ are feasible for detector noise below $\sigma \leq 0.02-0.03$ (Figure~\ref{fig:class-gauss} in Appendix~\ref{app:snr:gaussian-ests}).
We calculate that these $S/N$ levels can be readily achieved in a realistic confocal imaging system such as the one in \cite{Koklu2010} based on the range of powers that can be achieved with standard commercial lasers, diodes or power spectral densities offered by near-IR supercontinuum sources \cite{supercont}.

Cross talk between the signals coming from different gates presents a much more significant source of noise.
As shown in Figures~\ref{fig:gate_histograms}a,b this results from both the illumination spot driving polarizations along the wires both within and outside the gate of interest (Figure~\ref{fig:gate_histograms}a) as well as, at the detector plane, signal from the point spread functions of these current sources falling within the detection area (see, e.g. equations~\ref{eq:image_fields_psf} and \ref{eq:image_response_pixels}).

To examine these effects in detail we generated a series of tiles like the one shown in Figure~\ref{fig:image-concept}a and simulated their spectral images assuming the $0.8 \,\NA$ objective.
We simulated 20 images, each containing either 16 or 20 gates that were randomly selected from the Nangate library set.
This yielded on average $\sim 60$ observations, $S(c; \lambda, q)/S_0(\lambda, q)$, of each gate type with different surrounding gates.
\fix{The fluctuations in these observations therefore is due to the error term $\eta_{\delta}(c;\lambda, q)$ described in equation~\ref{eq:refl_to_ps}}.
The data for the XOR and AND gates under $y$-polarized illumination are shown in Figure~\ref{fig:gate_histograms}c,d as examples.
The series of curves follow similar paths to the net reflectance calculations presented in Figure~\ref{fig:gate_spectra}, but with reduced reflectance amplitude as a consequence of the lower collection efficiency of the $0.8 \, \NA$ that the image simulation accounts for.
The signals from the other gates and at $x$-polarization yield similar results.
We added a series of random Gaussian noise with $\sigma = 0.01$ to each of the simulated signal curves to account for detector noise, $\eta_{det}$ and used the resulting data to build histograms describing each $p(M_i \vert c_j)$.
Examples for the XOR and AND gate under $y$-polarized illumination are shown in Figure~\ref{fig:gate_histograms}c-g.
The full set of data are presented in the Figure~\ref{fig:si-class-full-maps-NA8} in appendix A.
The spreads tended to be fairly Gaussian with standard deviation of between 2 and 8 \% depending on the wavelength and polarization.

\begin{figure}[htb]
  \centering
  \includegraphics[width = 0.75\textwidth]{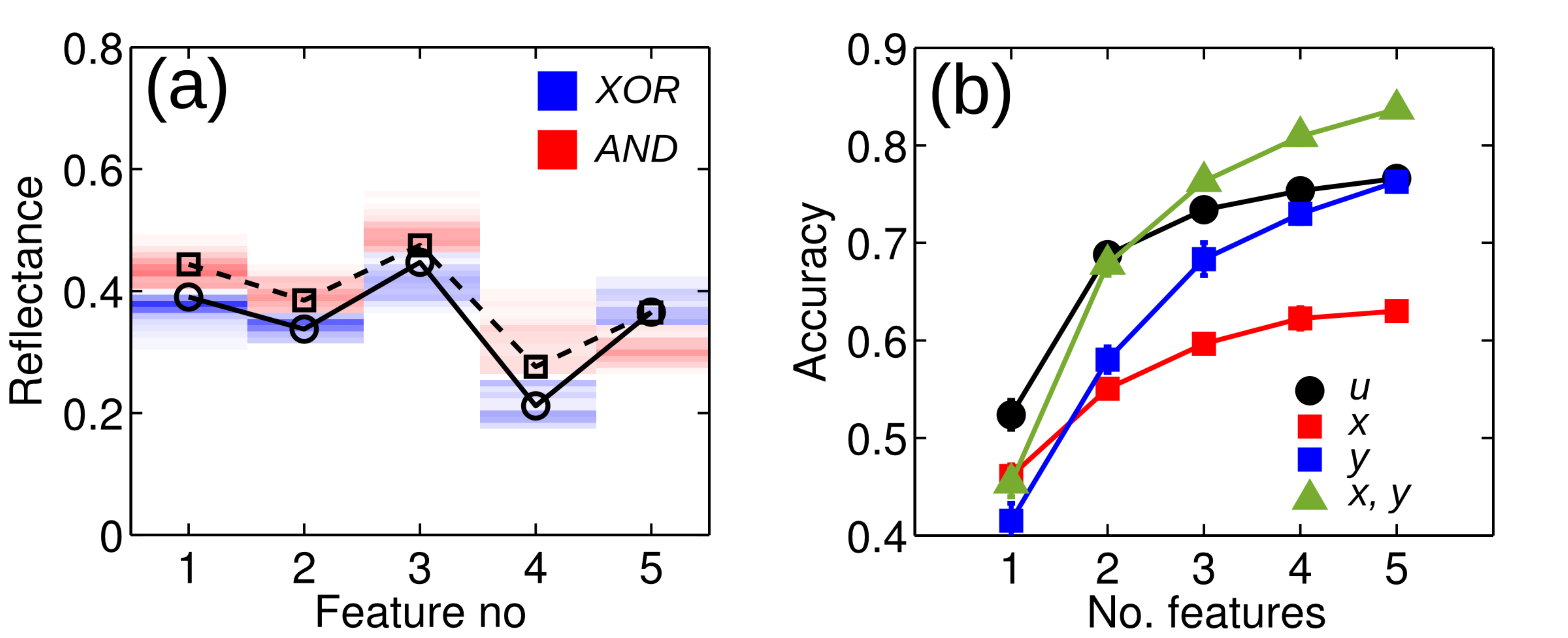}
  \caption{Bayesian classification for identifying gate spectral fingerprints. 
          (a) Overlaid probability maps for the XOR and AND gate probability maps at the spectral features selected. Taken with y-polarized light and an NA of 0.8. Darker shaded regions indicate higher probability of the gate.
          (d) Classification accuracy rates as a function of \fix{number of features}, for different options for the illumination polarization ($u$ - un-polarized, only $x$, $y$ - polarized or $x/y$ - both $x$ and $y$ available).} 
  \label{fig:classification_example}
\end{figure}

\section{Identifying Logic Gates}
\subsection{Classification Accuracies}
In order to characterize the overall accuracy of our approach while accounting for these two sources of noise, we used the large number of observations we obtained to \fix{train our classifier and} determine the error rates empirically.
\fix{
We selected (at random) 2/3 of all observations to use as a training set and reserved the other 1/3 to test the accuracy.
We used the training data to re-generate the histograms describing $p(M_i \vert c_j)$ as before (those in Figure~\ref{fig:gate_histograms} are based on the full set).
For a given set of measurements, $\vec{M}$, these distributions can be used to calculate the corresponding $P(M_i \vert c_j)$ and ultimately the probability of a given gate $P(c \vert \vec{M})$ via equation~\ref{eq:bayes} using the statistical independence simplification.
We used the test data to calculate the error rate for a given $\vec{M}$ by classifying each test observation and checking for accuracy such that the error rate could be determined empirically simply as the fraction of incorrect predictions.

While the feature space - wavelengths between $1$ and $3\, \mathrm{\mu m}$ and various polarization states - defines a large number of possible measurements, we aimed to select a small optimal subset that could readily be measured in a standard confocal microscope.
We first reduced the feature space by restricting the possible polarization states that could be selected according to 4 simple regularly encountered configurations.
These corresponded to unpolarized light ($q = u$), $q = x$, $q = y$ polarizations and the case where \emph{either} $x$ or $y$ polarization could be selected, i.e. $q \in {x, y}$.
The $q = u$ observations were computed as the average between the $q = x$ and $q = y$ signals.
This corresponds to the incoherent sum of the two signals as is representative of unpolarized (or natural) light \cite{Hecht2002}.
}
For each of these configurations we determined a set of 5 spectral features, $(\lambda_i, q_i)$, to use for measurements by applying a greedy algorithm that incrementally added to $\vec{M}$ by choosing the feature that resulted in the minimum error at each step.

A resultant set of spectral features and their probability distributions for the XOR and AND gate are shown in Figure~\ref{fig:classification_example}a.
Two measurements, for an XOR and AND gate, are shown as the black lines and markers.
Each measurement $\vec{M}$ traces out a path such that the probability of it belonging to a given gate class is determined from the underlying distribution.
Although the figure only shows the process for the two gates, all calculations considered the full set of 6.

We repeated the process of selecting features and calculating the overall error 10 times \fix{to cross validate our results}.
The resulting mean classification accuracies as a function of number of features measured are shown in Figure~\ref{fig:classification_example}b.
Error bars are included and correspond to $\pm$ one standard deviation, though are generally smaller than the markers.
The small deviation in conjunction with the fact that the set of features selected varied negligibly \fix{(see Figure~\ref{fig:si-features})} over each of the 10 runs indicates that an adequate sized data set was used \fix{to accurately represent the distributions (see Appendix~\ref{app:meth:class}).}
The results indicate that classification accuracies above 85\% are possible with only $N = 5$ features for the variable polarization case and nearly 80\% can be achieved even without polarization control.  
This is a significant improvement over the $16\%$ random baseline.

\subsection{Spatial sampling}

\begin{figure}[htb]
  \centering
  \includegraphics[width=\textwidth]{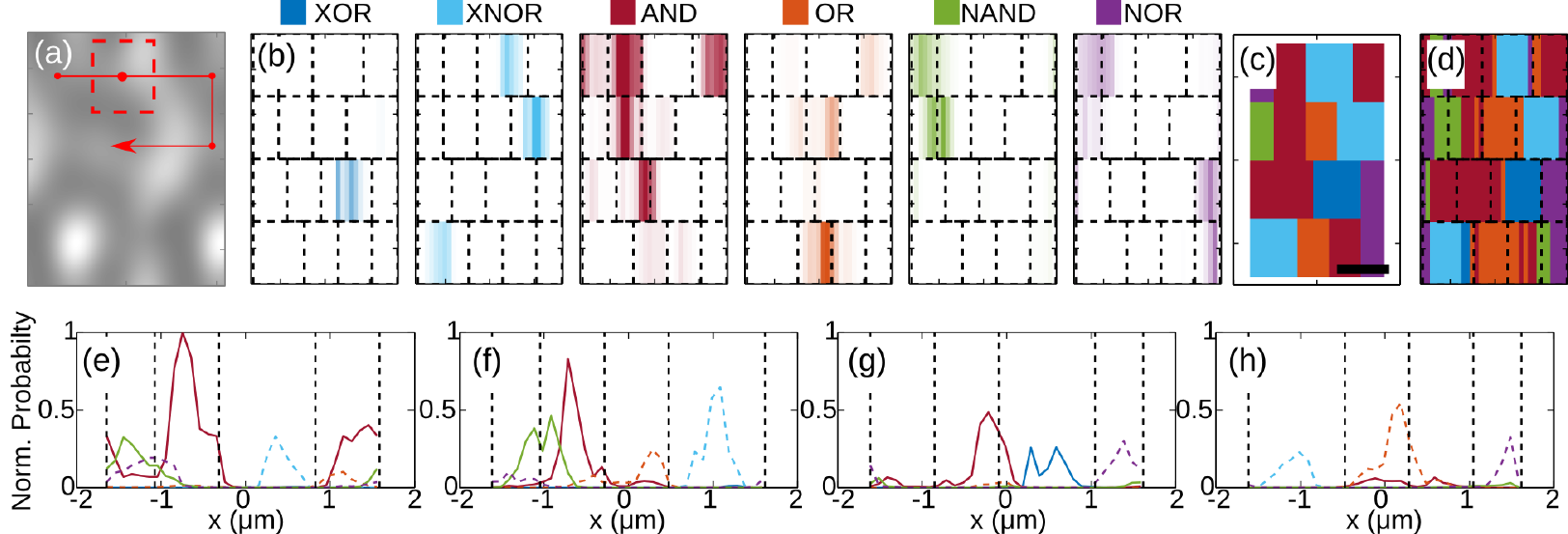}
  \caption{Gate probability maps and reconstructing the gate level map. 
          (a) Illustration of the spatial sampling used in the calculation. The background shows the raw simulated intensity at the detector plane. The integration window and its path across the image are illustrated via the red lines. 
          (b) Spatial maps with each pixel color indicating the probability of a given gate \fix{(i.e. $P(c \vert \vec{M})$)}
          (c) True (object) spatial map of the gates for the test object. 
          (d) Final gate map determined by assigning each pixel to the gate associated with the maximum probability.
          (e-h) Cuts taken along each row showing the probability of a given gate from top to bottom.
          } 
  \label{fig:gate_map}
\end{figure}

The results in Figure~\ref{fig:classification_example} correspond to the ideal case for classification where each pixel is perfectly centered at each gate's location.
In principle this can be achieved if the gate dimensions in the $x$-direction share a common factor.
This turns out to be the case for the Nangate library we used here, but this may not always be true depending on the IC library used and effects of sampling and mis-alignment are therefore important to address.
In order to gain greater insight into these issues we generated a series of maps from a test image, using the previously described training data and feature set.
The results for a high-sampling rate ($\Delta = 100 \units{nm}$) along $x$ are shown in Figure~\ref{fig:gate_map}.

Figure~\ref{fig:gate_map}b shows the spatial variation of $P(c \vert \vec{M})$ for each gate (darker color corresponding to higher probability).
In comparison with the true gate map shown in Figure~\ref{fig:gate_map}c the agreement is excellent.
Displaying the probabilities themselves provides greater insight into the sources of error and effects of mis-alignment.
In particular, the confusion between the similar pairs of gates, such as (XOR, XNOR), anticipated based on Figure~\ref{fig:gate_spectra} is immediately confirmed, especially for both the (AND, OR) and (NAND, NOR) pairs.
The correct probabilities in general tend to peak near the center of each gate and also maintain relatively high values (in comparison the the incorrect gates) over the majority of the pixels falling within the gate (Figures~\ref{fig:gate_map}e-h).
This indicates that the classification should be robust to misalignments and also enable accurate determination of the gates' centers.
The final map, with each pixel classified according to the maximally probable gate, is shown in d.
The image closely matches the true layout, with errors occurring primarily at the boundaries between two different gates as expected.

We generated a series of probability maps and gate maps for different sampling rates in the $x$ direction, raging from $100$, $250$ and $500$.
The full results are shown together in Figure~\ref{fig:si-spatial-sampling} in the appendix, and the gate map sampled every $250$ nm is also reproduced in Figure~\ref{fig:image-concept}.
In all cases similar excellent agreement with the ground truth gate distribution was obtained.
Quantitatively, in each case approximately 70\% of the pixels were assigned to the correct gate and between 80 and 85\% were assigned to one of the correct group (i.e. XOR or XNOR; AND or OR; NAND or NOR).
Our results indicate that with minimal alignment (i.e. only in the vertical dimension) and no a-priori information about the underlying gate layout an accurate map can be obtained.
The robustness of the classification to pixel alignment points to the ability to down-sample significantly without sacrificing accuracy, as long as each gate can be sampled.

\section{Conclusions}\label{conclusions}
Our results demonstrate that backside spectral imaging of ICs in conjunction with established pattern recognition techniques provides a powerful route towards probing their functionality.
The method we propose addresses key obstacles associated with optical resolution and sampling that arise due to the nanoscale features size and massive complexity of modern ICs.
These challenges fundamentally limit the ability of traditional imaging techniques to rapidly map a large fraction of a chip.
The reduction in resolution requirements we achieve is particularly significant since it overcomes a fundamental physical limitation.
ITRS estimates of the trends in gate sizes \cite{ITRS} imply that our approach can easily scale well beyond the nominal "10 nm" nodes, where any direct imaging method is completely inconceivable.

Our method has immediate applicability to the significant problem of ensuring the security of ICs.
To our knowledge, no other technique offers as direct and informative a measure of the physical layout of a fabricated IC in a rapid and non-destructive fashion.
The information we obtain can be used to directly verify the IC layout and spatially localize suspicious regions.
\fix{
We note that the accuracy in detecting actual Trojan circuits will be different, though we expect higher, than the single gate prediction accuracies we have demonstrated here. 
This is because mis-classifications of the type that constitute actual errors will appear as random noise dispersed throughout the layout while additional, modified circuitry will often be localized \cite{Nowroz2014} and have structure defined by electronic functionality, routing and floor planning requirements \cite{Chakraborty2009,Imeson2013}.
Combining the data we generate with other testing methods, such as those that serve to constrain easily modified dead space in the IC layout\cite{Zhou2015} should enable very efficient, accurate and informative testing.
}

\fix{
Additionally, many possibilities exist for improving the single pixel accuracy from the standpoint of both computational pattern recognition as well as optical design.
Alternative more advanced pattern recognition methods \cite{Baker2014} in conjunction with leveraging additional a-priori information can provide significant improvements.
For example, the physical dimensions of the gates are known and can be taken into account when generating the maps such as those shown in Figure 6 to avoid the isolated mis-classified pixels that are intuitively infeasible.
From the standpoint of optical design, optimizing aspects such as $\NA$ and pinhole size may also offer improvements.
Determining ideal parameters for a given gate size is an interesting question, since it involves a variety of trade-offs such as maximizing the diversity of the gates' spectral response, robustness to mis-alignment and minimizing cross-talk.
A particularly interesting configuration would be to illuminate at above the critical angle from the Si side, such that an evanescent wave can be used to excite the M1 layer similar to the TIRF measurements that are widely used in biological applications.
This offers a means to significantly reduce signal from the metal interconnect layers above M1 that may produce additional background noise that we were not able to consider here.
}

\fix{We note that the concepts we have introduced here are extensible to other IC applications beyond gate substitutions in security.
 In general,} the sensitivity of the spectral scattering response to nanoscale structure points to the opportunity to obtain more detailed information.
For example, it is well known that forming conductive bridges in nanoscale metallic structures dramatically alters their antenna-like characteristics.
This provides a clear route to easily detecting bridging defects which are becoming increasingly significant as a result of increased wire density \cite{Ma2012,Goel2009}.
This is relevant to failure analysis where related concepts of leveraging prior information to improve spatial resolution are beginning to be explored \cite{Cilingiroglu2015,Cilingiroglu2015b}.

In general, ICs represent a compelling and ideal application of the spectral fingerprinting methods we employ here.
They are inherently modular since their design paradigm is based on abstraction, hence their components fit neatly into a few well defined classes.
Their physical structures are also highly uniform - an enormous amount of effort is applied to ensure this on the fabrication end in order to provide adequate yield.
This implies that in physical measurements sample variations will be minimized.
Both of these are dramatic departures and improvements over the situations where spectroscopic imaging techniques are often applied such as in biological systems.
Access to sufficient and representative samples to use as training data is a final critical need in classification techniques that ICs offer perhaps better than any other physical system.
A single IC contains typically over a million gates that are repeated and reused throughout \emph{multiple chips on multiple fabrication runs}.
Measurements on a small fraction of a single chip can supply training data that can be used in a classification method that will be applied to millions of chips.
Spectroscopic imaging techniques therefore have the potential to satisfy critical metrology needs, in security and also general applications, that are increasingly significant as the feature sizes and complexity in modern ICs passes beyond what can be addressed with traditional tools.


\appendix
\section*{Appendix}
\section{Methods}
\renewcommand{\theequation}{A\arabic{equation}}
\setcounter{equation}{0}
\label{app:meth}

\subsection{Spectral Imaging in a Scanning Confocal Microscope}
\label{app:meth:image-sig}
The calculations and considerations we presented are representative of a scanning confocal microscope.
While the general functionality and correspondence is described in the main text, we derive here all equations used and define their correspondence to a practical microscopy setup.
As the central SIL configuration exactly corresponds to our computational results and is the simplest, we assume this throughout.
For simplicity, we consider only the wavelength dependence of components and the sample.
The extension to include polarization, $q$, is straightforward and essentially amounts to the substitution $\lambda \rightarrow \lambda, q$ for any elements that are polarization dependent.

A scanning confocal microscope illuminates the sample with a focused spot and records the reflected power that passes through a detector plane aperture.
In a practical microscope, the power delivered to the focal region with the illuminating beam can be written as,

\begin{equation}
  P_{foc}(\lambda) = \left[1 - \abs_{Si}(\lambda) \right]\trans_{Si}(\lambda)\mu_f(\lambda)P_0(\lambda)
\end{equation}

\noindent where $P_0(\lambda)$ is the source power.
This equation accounts for loss in the various free space optical components, transmission through the Si-Air interface at the surface of the SIL and absorption in the Si substrate via the parameters $\mu_f$, $\trans_{Si}$ and $\abs_{Si}$ respectively.
While the beam focused through the SIL is comprised of plane waves propagating at a range of angles (i.e. its angular spectrum), all are normally incident at the Si-Air interface in the central-SIL configuration (see Figure~\ref{fig:image-geometry}) thus $\trans_{Si}$ is solely a function of wavelength.
This is also true, to good approximation, for absorption in the substrate ($\abs_{Si}$) since for all but the highest $\NA$ systems the difference between the path lengths of the rays associated with the minimum and maximum angles is negligible.
For the $0.8\NA$ system we focus on here the maximum angle in the Si is about $13$ degrees, which leads to an about 3\% increase in path length versus the 0 angle ray.
Hence $\abs_{Si}(\lambda) = \exp(-\beta(\lambda) d)$, where $\beta(\lambda)$ is the absorption coefficient of Si and $d$ is the substrate thickness.
Upon interacting with the sample a fraction of the incident power will be reflected.
The total power reflected encompasses all propagating rays scattered into the substrate side,

\begin{equation}\label{eq:pwr_refl_tot}
  P_{\refl}(x, y; \lambda) = \refl_{S}(x, y;\lambda)P_{foc}(\lambda)
\end{equation}

\noindent The quantity $\refl_S(x, y; \lambda)$ is the net reflectance of the sample - the fraction of the total incident power that is reflected.
Generally some fraction, $f_C$, of the total reflected power is collected due to the finite cone of angles that can be captured by the collection objective.

Accounting for this fraction, and propagation back out through the Si, SIL and detection optics, the total power incident on the detector plane is,

\begin{align}\label{eq:pwr_refl_det}
  P_{D}(x, y; \lambda) &= \mu_{c}(\lambda)\trans_{Si}(\lambda)\left[1 - \abs_{Si}(\lambda)\right]^2\refl_{S,C}(x, y;\lambda)P_{foc}(\lambda) \\
                 &= \mu_{c}(\lambda)\left[ \trans_{Si}(\lambda)\right]^2\left[1 - \abs_{Si}(\lambda)\right]^2\refl_{S,C}(x, y;\lambda)\mu_f(\lambda)P_0(\lambda)
\end{align}

\noindent where $\refl_{S,C} = f_C\refl_S(x, y; \lambda)$ is the fraction of the input power reflected into the collection cone of the system and $\mu_c$ describes loss due to practical collection optics.

The signal captured by the detector, described in equation~\ref{eq:image_response_pixels}, can therefore be written as,

\begin{align}\label{eq:det_sig_ab_2}
  S(a,b) = \alpha(\lambda)f_\mathcal{W}(x, y;\lambda)P_D(x, y; \lambda)
\end{align}

\noindent where the quantity $f_\mathcal{W}(x, y;\lambda)$ gives the fraction of the total power incident on the detector that is contained within the area defined by the aperture (i.e. confocal pinhole).
The value depends not only on the geometry of the pinhole and the point-spread function of the imaging system, but also on the particular sample geometry.
For example, even in the case of a beam reflected from a planar interface, a metallic surface will produce a tight spot at the detector plane while a high-to-low refractive index interface will produce a significantly larger spot that carries much of its power in a series of side-lobes.
We associated the position independent factors with a system response, $R_D = \alpha\mu_c\trans_{Si}^2(1-\abs_{Si})^2\mu_f$.

In a scanning confocal microscope the image is formed pixel-by-pixel by translating the sample, either physically or effectively, through the use of galvo-mirrors and a relay, and recording the signal 

\begin{align}\label{eq:det_sig_scan}
  S(a_i, b_j;\lambda) &= R_D(\lambda)f_\mathcal{W}(x_i, y_j)\refl_{S,C}(x_i, y_i; \lambda)P_0(\lambda)
\end{align}

\noindent assuming regular sampling intervals in the $x$ and $y$ directions, $(x_i, y_j) = (x_0 + i\Delta_x, y_0 + j\Delta_y)$.

Non-sample dependent contributions to the signal can be removed by normalizing by a suitable background, ideally a mirror at the sample plane.
In this case, the background signal is just $S_0 = R_Df_{\mathcal{W},m}P_0(\lambda)$ since $\refl_{S,C}(\lambda) = 1$ for an ideal mirror.
Thus,

\begin{align}\label{eq:det_sig_rf0}
  S(a_i, b_j; \lambda)/S_0 &= f_\mathcal{W}(x_i, y_j; \lambda)\refl_{S,C}(x_i, y_j;\lambda)/f_{\mathcal{W},m}(\lambda) \\
                  &= \refl(x_i, y_j; \lambda)
\end{align}

where the generalized reflectance $\refl(x_i, y_j; \lambda)$ is still a ratio of power out versus input, but includes also effects due to the detector plane aperture.
This is commonly the response that is recorded in microspectroscopy experiments~\cite{Davis2010}
If we account for detector noise then $S(a_i, b_j; \lambda)/S_0 = \refl(x_i, y_j; \lambda) + \eta_{det}$.

The reflectance of a given gate class $c$, $\refl(c; \lambda)$, is given by $\refl(x_i, y_j; \lambda)$ when $(x_i, y_j)$ correspond to the center coordinates of the gate.
Therefore,

\begin{align}\label{eq:det_sig_rff}
  S(a_i, b_j; \lambda)/S_0 &= \refl(c;\lambda,q) + \eta_{det} + \eta_{\delta}
\end{align}

\noindent where $\eta_{\delta}$ is the noise due to the interference of the gates surrounding the central gate. Equation~\ref{eq:det_sig_rff} gives equation~\ref{eq:refl_to_ps} in the main text.

\subsection{FDTD Simulations}
\label{app:meth:fdtd}
We used a commercial FDTD solver to simulate the response of the IC M1 lines and contact layers for all calculations in the main text. 
All simulation geometries consisted of a semi-infinite Si substrate occupying the lower half-space and a $\mathrm{SiO_2}$ background in the upper-half.
Both were treated as non-dispersive, loss-less materials with refractive indices of $n_{Si} = 3.5$ and $n_{SiO2} = 1.45$.
The M1 layer and contacts were copper and tungsten, respectively, both treated as dispersive material with realistic relative permittivities taken from fits to the data in \cite{Palik1988}.
We neglected the active regions and polysilicon transistor gates for simplicity and due to the fact that they are expected to contribute significantly less to the gate signals in comparison with the metal structures.
The M1 layer and contacts were copper and tungsten, respectively, and embedded in a homogeneous $\mathrm{SiO_2}$ ($n = 1.45$) dielectric background, with the M1 situated a height of 100 nm above the Si substrate.
The height of the M1 layer was 130 nm.
All lateral dimensions were taken from the Nangate library standard cells~\cite{Nangate45}.

The calculations in Figure~\ref{fig:gate_spectra} of the main text assume illumination with a uniform, linearly polarized, normally incident plane wave.
The boundary conditions are periodic in the lateral plane and absorbing (PMLs) in the $z$ direction.
Each gate simulated defined the unit cell such that the results correspond to an infinite tiling in the lateral plane.
The power transmitted through a plane in the Si lower half-space, behind the source, was calculated and normalized to the source power to determine the net reflectance.

\subsection{FDTD Simulated Microscopy Images}
\label{app:meth:fdtd-image}
We simulated the images of the IC gate components by calculating the far-field angular spectrum, $\angy{E}_\infty(s_x, s_y)$, from the FDTD simulations.
The near-fields recorded on a plane behind the source in the Si substrate were used to compute the far-fields via the near-to-far-field transform \cite{Lumerical}.
The Si - $\mathrm{SiO_2}$ interface is automatically accounted for since it lies within the simulation domain.

The angular spectrum represents the far-field as a series of plane-waves defined by the direction cosines $s_x = \sin\theta\cos\phi$ and $s_y = \sin\theta\sin\phi$, where $\theta$ and $\phi$ are the usual spherical coordinates (see Figure 1).
The far-fields computed from the simulations are those inside the Si substrate, and must be propagated through it, and the imaging system consisting of the collection objective and tube-lens.
\fix{
Neglecting refraction at the Si-Air interface (as with, e.g. a central-SIL), the relative weighting of the plane wave components stays constant despite overall reductions in amplitude due to absorption in the Si substrate and reflection at the interface. We account for the reduction in overall power incident at the detector and its implications for S/N in the relevant calculations (see Appendix A1 and Appendix B1).
}
Therefore, the image at the detector plane can be formed via,

\begin{align} \label{eq:full-E-det}
  \vec{E}_D(u, v, w) &= \frac{jn_1\sqrt{n_1}}{\lambda M} \iint_{\Omega_{\NA}}
               \angy{E}_\infty(s_x, s_y)e^{jk_0n_1(s_xu/M + s_yv/M)}\, 
               d\Omega
\end{align}

\noindent where the integral is over the collection cone of the objective $\NA$ \cite{Richards1959,Davis2010,Davis2010b}.

Equation~\ref{eq:full-E-det} can be used directly to compute the image formed by a confocal imaging system where the object is illuminated by a finite sized beam (in principle with the same $\NA$ as the collection objective) and the object is simulated as isolated via standard $PML$ boundary conditions.

Because the full confocal simulation approach requires a focused beam excitation source a fairly large computational domain is required and each wavelength point requires a separate simulation.
The latter requirement is due to the fact that the source is modeled as a sum of plane-waves that are defined by a fixed $\vec{k}$ vector such that the angle of incidence changes with wavelength \cite{Liberman2011}.
In order to reduce the computational burden when simulating the large number of IC gate images over a broad spectral bandwidth that was required to train and test our classifier we performed a simplified calculation based on two approximations: (i) simulating the object as periodic \cite{Davis2010b} and; (ii) exciting the simulation with a normally incident uniform plane-wave.
In terms of the far-field, the resultant angular spectrum is a discrete series corresponding to the diffraction orders,

\begin{equation} \label{eq:discrete}
  \vec{E}_D(u, v, w_0) = \frac{1}{M}\sum_{k,l \in \Omega_{NA}}\vec{s}_{k,l}e^{j(kG_xu/M + lG_yv/M)}
\end{equation}

\noindent where $G_x = 2\pi/(p_x)$, $G_y = 2\pi/(p_y)$ are the grating wavevector magnitudes with $p_x$ and $p_y$ corresponding to the $x$ and $y$ dimensions of the unit cell.
The weightings $\vec{s}_{k,l}$ are normalized such that $\vert \vec{s}_{k,l} \vert^2$ give the relative fraction of the incident power in each grating order \cite{Lumerical} and $|\vec{E}_D|^2$ from equation~\ref{eq:discrete} gives the power for unit area.
Because the source has infinite extent over the full simulation domain, to calculate the reflectance of each pixel in the multi-gate simulations we normalized the integrated intensity over the integration window $\mathcal{W}$ to the power delivered to the same area in the object space.

This approach dramatically reduces the required number of simulations.
It enabled us to simulate a large unit cell, which corresponded to a 4x4 or 5x4 tiling of gates, at once and record all wavelengths with a single simulation.
Hence we were able to generate the over 350 total gate observations each of which contained 50 wavelength points with 20 FDTD simulations.
In contrast, without the approximation one simulation is required per gate observation per wavelength point such that 17,500 simulations would have been required.

The cost of the approximations used are that they neglect any illumination side effects or spatial selectivity and therefore over-emphasize cross-talk between adjacent objects that are approximately a point-spread-function distance apart.
This is because all objects are excited coherently and uniformly.
In the case of illumination by a focused beam, only objects within the illuminating spot are excited.
For objects that are much closer together than the size of the focused spot, the approximation is not significant since they will be excited coherently with the same amplitude in both cases.
Objects that are much further apart than the point-spread-function of the collection objective, will be incorrectly excited coherently, but the spots they produce on the detector plane will not overlap.
The 0.8 NA with which we performed all calculations yields a spot size on the order of $1 - 2\, \mu \mathrm{m}$ in the near-IR, which is roughly on the order of the gate size.
The wires in the gate are therefore driven uniformly, but neighboring gates are excited to a lesser degree.

\subsection{Gate Identification}
\label{app:meth:class}
\fix{We empirically calculated the error rate of our classifier by segmenting our data into a train and test set and using the latter to determine the rate of mis-classifications.
If insufficient data is present niether the train nor test data sets will accurately represent the true statistical behavior of the observations.
As a result the features selected and error rates will depend strongly on the test and train data.
If suffient data are present, not matter how the observations are divvied up, the statistical distributions they describe will be the same.
Therefore both the error rate and features selected should be consistent irregardless of exactly which observations are put into the test or train set.
}

\begin{figure}[h!tb]
  \includegraphics[width = \textwidth]{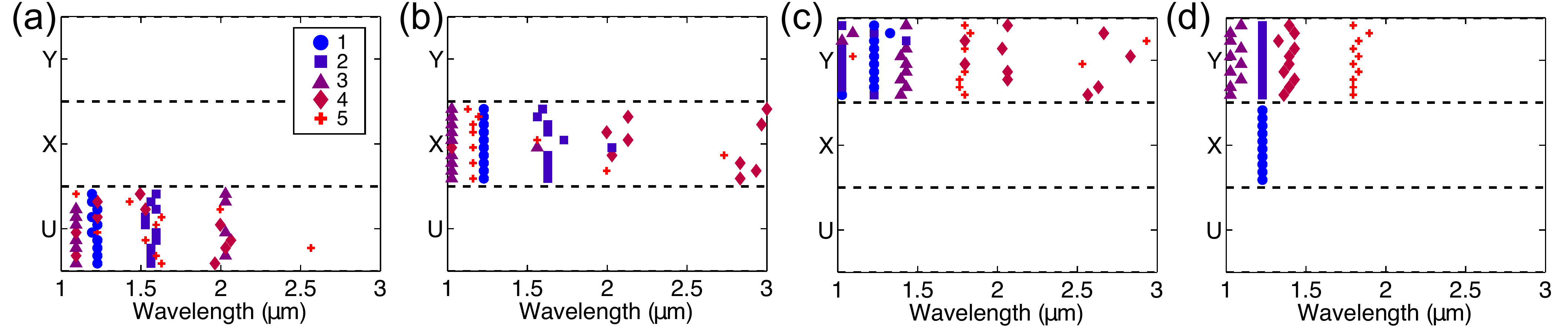}
  \caption{Feature selection over multiple runs.
       The features selected during the empirical feature selection and error calculation process are shown for each of the four polarization cases (a - $u$), (b - $x$), (c - $y$) and (d - $x,\, y$).
       Each row in the panels represents one of the 10 runs.
       The different markers indicate the order the features were selected in, the first, $1$, being the mosti important and last, $5$th, being the least.}
  \label{fig:si-features}
\end{figure}

\fix{
To verify that this was the case here, we randomly segmented our data into test and train sets and determined features and error rates.
We repeated this process 10 times.
As shown in Figure~\ref{fig:classification_example}b in the main text, the error rate varied negligibly
}
Figure~\ref{fig:si-features} shows the features chosen to classify the gates for each of the four polarization cases.
The features selected are \fix{also} extremely consistent across all 10 runs.
Features selected earlier were highly consistent.
This is expected since they are the most useful in discriminating between the different gates.
Features selected later tend to be less consistent as they offer less improvement.
This is reflected in the slopes of the accuracy curves in Figure 5 of the main text.

\subsection{Integrated Circuits and Simulated Training Samples}
\label{app:meth:gates}
All integrated circuit gate layouts are taken from the Nangate 45 nm open source library \cite{Nangate45}.
The library specifies the planar geometries of the various layers - diffusion, polysilicon, contact, metal 1 - present in each gate.
Throughout, for simplicity, we limited our analysis to the metal 1 and contact layers and Si substrate.
Due to the lower refractive index contrast the diffusion and polysilicon layers will contribute to the signal to a significantly smaller degree.

The specific gates we selected for our test set were the six smallest 2-input fundamental gates in the library.
These are denoted, in full, `XOR2\_X1', `XNOR2\_X1', `AND2\_X1' and so on for the other gates.
All gates in the set, and in the entire library, are consistently dimensioned at $1.14 \,\mu\mathrm{m}$ in the vertical direction.
This is due to the fact that in an IC the source voltage and ground rails are arranged on a regular 1D grid and the gates tiled in between them (see e.g. Figure 2).
The horizontal dimensions are $d_i = 1.14 \,\mu\mathrm{m}$ for the (XOR, XNOR), $d_j = 0.76 \,\mu\mathrm{m}$ for (AND, OR) and $d_k = 0.57 \,\mu\mathrm{m}$ for (NAND, NOR).

\begin{figure}[h!tb]
  \includegraphics[width = \textwidth]{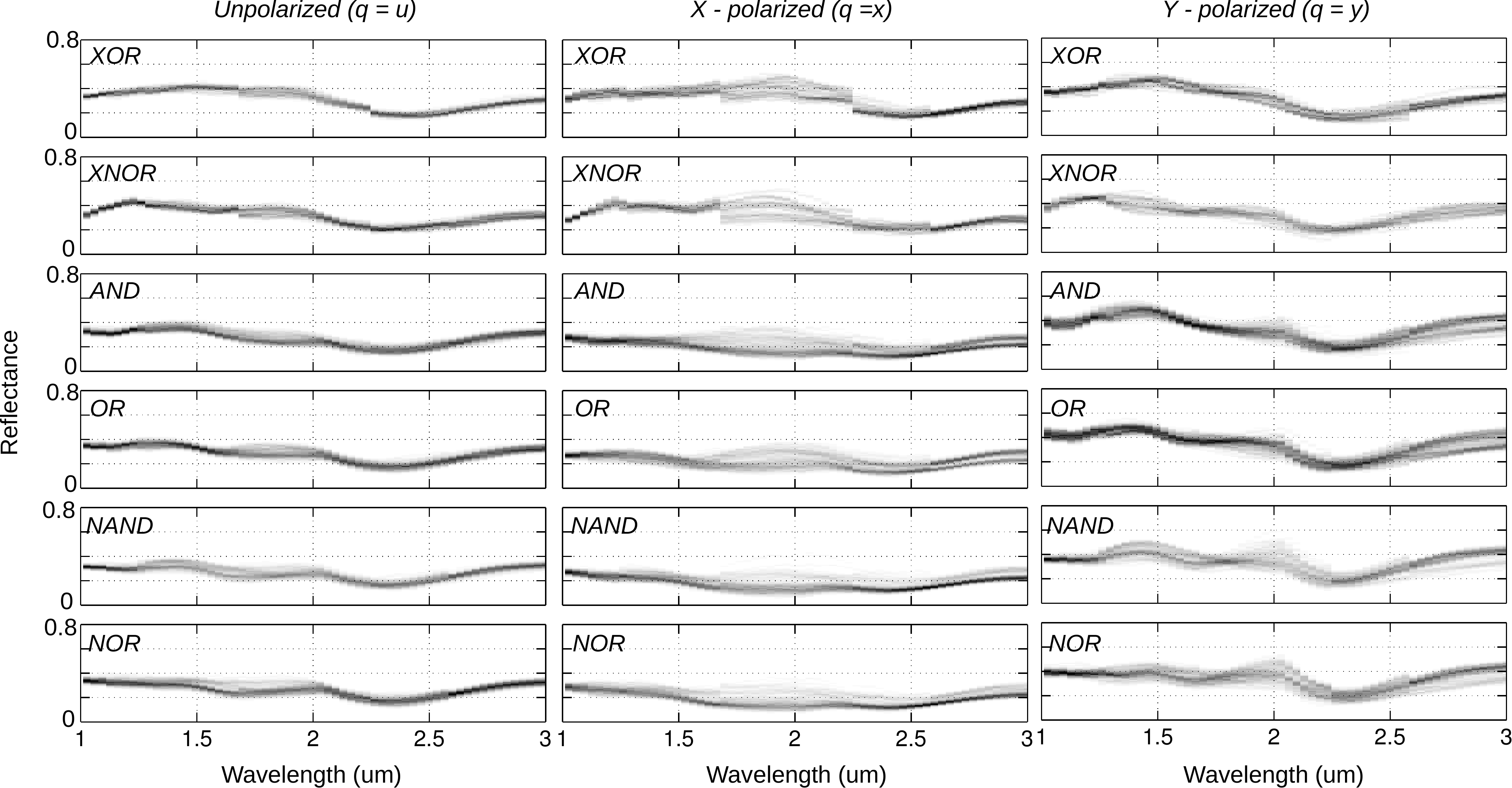}
  \caption{Spectral probability maps for the six fundamental logic gates imaged using a $0.8\NA$ collection objective. 
  The maps are shown for unpolarized (left) $x$ (middle) and $y$ (right) polarized illumination.
  The type of each logic gate is indicated. The apparent 'jumps' in some of the spectra are due to diffraction orders and are artifacts of the periodic simulation.}
  \label{fig:si-class-full-maps-NA8}
\end{figure}

\begin{figure}[htb]
  \centering
  \includegraphics[width = 0.75\textwidth]{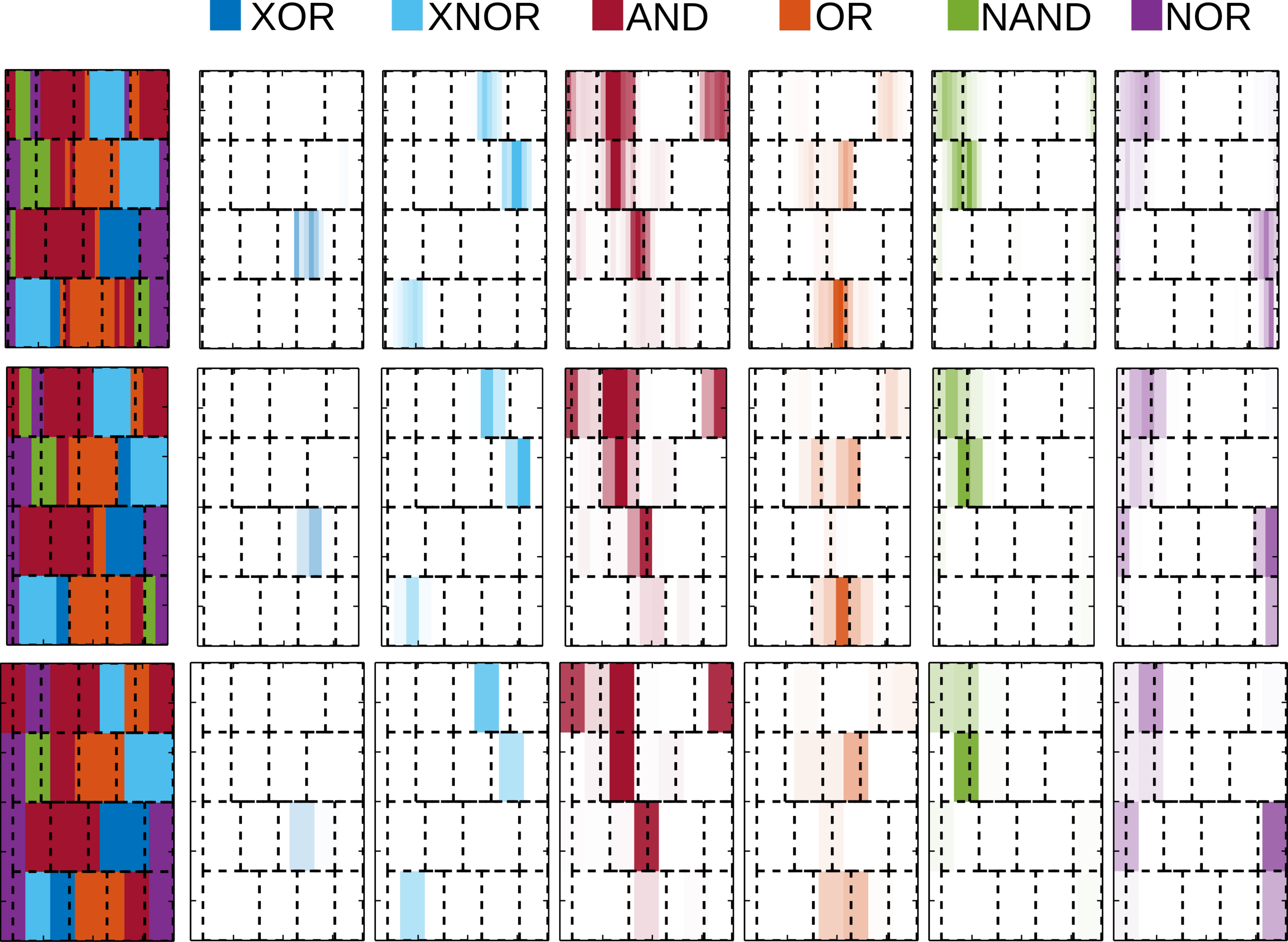}
  \caption{Influence of spatial sampling on the acquired probability map.
       Spatial gate and probability maps are generated as described in the main text.
       The spatial sampling interval in the $x$ direction is 100, 250 and 500 nm from the first to bottom row.
       The first column shows the gate level map with each pixel assigned to a gate class via the Bayes' decision rule used throughout.}
  \label{fig:si-spatial-sampling}
\end{figure}

To generate the various spectral responses and histograms shown in Figure 4 in the main text and Figure~\ref{fig:si-class-full-maps-NA8} we generated a series of random tilings of these gates and simulated their images.
The gates are arranged on a rectangular grid in the $y$ direction in a series of rows as illustrated in Figure~\ref{fig:image-concept} of the main text.
The gates are placed randomly, subject to the constraint that all the rows must have a fixed length in $x$ to be compatible with the periodic unit cell of our simulation approach.
We chose a unit cell size that corresponded to 4 rows and 4 gates per row with one of each pair type plus an addition gate of the middle size (total length $d_i + 2d_j + d_k$).
Because the gates' horizontal dimensions share a common factor, this unit cell could also be solved for with 5 gates per row (2 of type $j$ and 3 of type $k$).
For each new simulation one of the two possible solutions was selected at random (equal probability).
For each row in the tilings, gate order and the exact gate for each type (i.e. XOR or XNOR) was again selected randomly for maximum diversity.
In all we generated 20 different tilings which yielded a total of 22 observations each for the XOR and XNOR gates (the least frequent gates) and over 75 for each of the remaining gates in the set.
The resultant probability distributions for all gates and polarizations are shown in Figure~\ref{fig:si-class-full-maps-NA8}.
We additionally generated a new image, not used in any of the feature selection or error calculations, to generate the results in Figures 6 and \ref{fig:si-spatial-sampling}.
Our spatial classification example described in Figure 6 of the main text is shown for 3 different $x$ direction sampling rates in Figure~\ref{fig:si-spatial-sampling}.

\section{System and Detector Noise Effects}
\renewcommand{\theequation}{B\arabic{equation}}
\setcounter{equation}{0}
\label{app:snr}

\subsection{Signal-to-noise ration estimates}
\label{app:snr:pwr-budget}
In order to estimate the influence of detector noise we used the usual definition of signal-to-noise ratio ($S/N)$,

\begin{equation} \label{eq:snr}
  S/N = \frac{i_S}{i_N} = \frac{R(\lambda)P_S}{\sqrt{\Delta f}R(\lambda_p)P_{NEP}}
\end{equation}

where $i_N$ is the standard deviation in the detected signal, $R(\lambda)$ is the detector responsivity and $\lambda_p$ denotes the peak responsivity wavelength.
The signal power, $P_S$, depends on the sample reflectance via $P_S = \refl \beta P_0$, where $P_0$ is the source power and $\beta$ scales for loss in the measurements system.
Converting to a reflectance scale amounts to dividing all signals by $P_S(\refl = 1)$ such that the standard deviation in reflectance is given by $\sigma = (S/N)^{-1} = P_{NEP}\sqrt{\Delta f}/P_S$ as in the main text, where we have assumed $P_S$ is measured for a perfectly reflecting sample at the peak responsivity.
Assuming the noise is purely additive, this $\sigma$, can be used for all the probability distributions when approximated as Gaussian.

Commercially available extended InGaAs detectors have a $P_{NEP} \sim 10^{-12} \, \mathrm{W/\sqrt{Hz}}$ and a high responsivity between $\lambda = 1 - 2.6 \, \mu \mathrm{m}$ \cite{Thorlabs_InGaAs}.
From equation~\ref{eq:snr}, the required signal power to achieve a given $S/N$ is given by,

\begin{equation} \label{eq:PS_det}
  P_{S,det}(S/N, \lambda, \Delta f) = (S/N)\sqrt{\Delta f} R(\lambda_p)P_{NEP}/R(\lambda)
\end{equation}

\begin{figure}[htb]
  \centering
  \includegraphics[width= 0.75\textwidth]{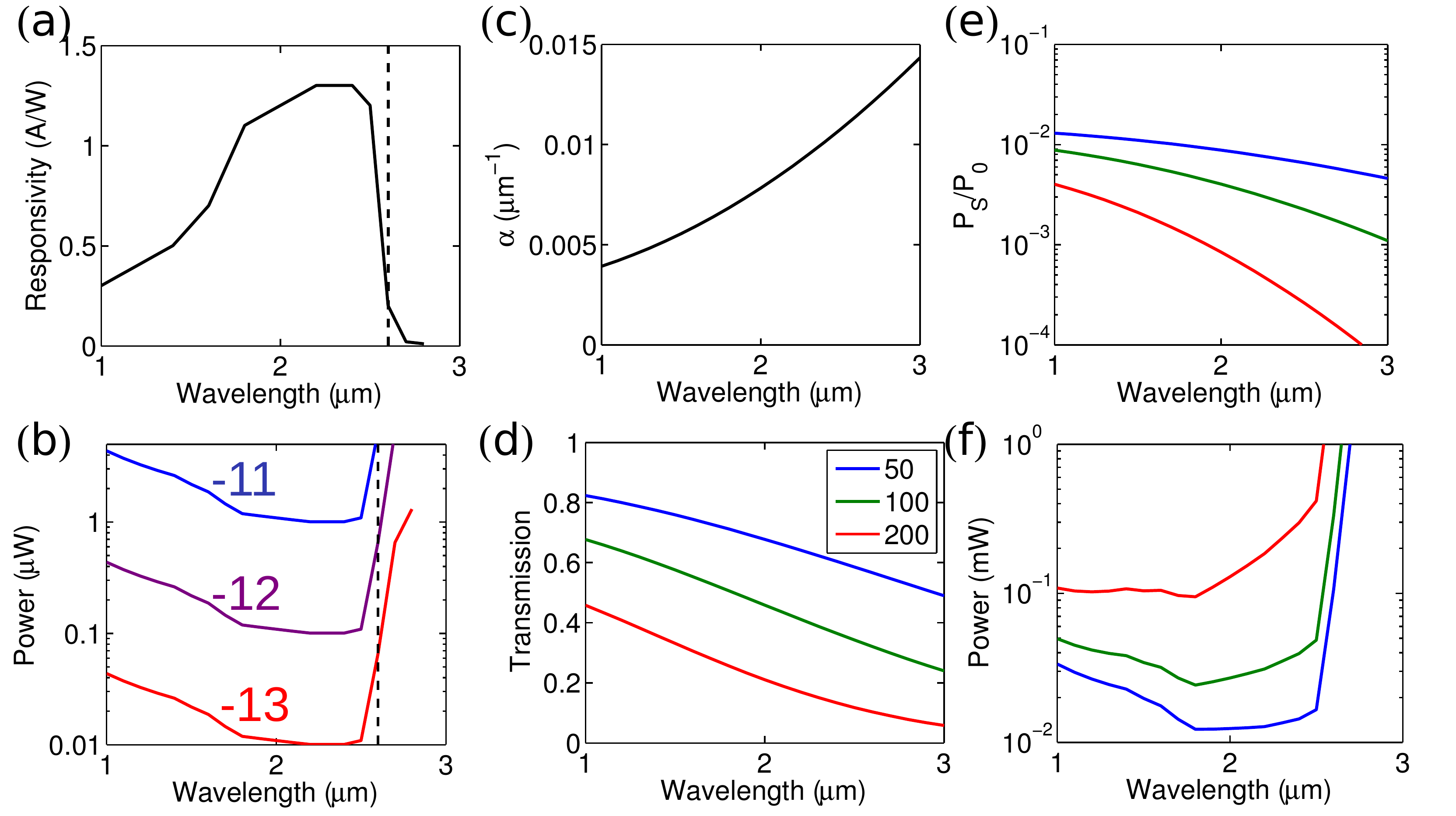}
  \caption{Required power for $S/N = 100$ measurements at $\Delta f = 1\units{MHz}$ and influence of optical system loss. 
  (a) Responsively of a Thorlabs extended InGaAs biased detector. The detector's $P_{NEP} = 1\times 10^{-12}$. 
  (b) Power required for different values of $P_{NEP}$ ($10^{-11}$, $10^{-12}$ and $10^{-13}$ as indicated assuming the InGaAs responsively curve in (a).
  (c) Absorption coefficient for a heavily doped Si substrate. Extracted from Fig. 1 of \cite{Paniccia1998}. 
  (d) Transmission (single pass) through 50, 100 and $200\, \mu \mathrm{m}$ thick substrates.
  (e) Fraction of input power received at the detector for three different thickness Si substrates. Same legend as in (d). 
  (f) Required input power from the source accounting for all sources of loss and detector responsivity.}
  \label{fig:Pnep}
\end{figure}

Figure ~\ref{fig:Pnep} gives a sense of the required powers to produce a $S/N$ of 100 at MHz imaging rates (i.e. $\Delta f = 1\units{MHz}$) for a typical commercial InGaAs detector, calculated according to eq.~\ref{eq:PS_det} where the responsivity curve is extracted from the data sheet in \cite{Thorlabs_InGaAs}. 
As stated in the main text, power levels on the order of $0.1 - 1\,\mu\mathrm{W}$ should be sufficient to achieve $S/N > 100$ therefore $\sigma \sim 0.01$.

To estimate the required \emph{input} power to generate the desired signal levels at the detector we need to account for the sources of loss in the optical system as outlined in Appendix A1.
Assuming $\mu = \mu_c \approx \mu_f$, and that the detector plane aperture is not excessively small, $P_D = \left[\mu \trans_{Si} (1 - \abs_{Si})\right]^2P_0$.
For a system like the one in \cite{Koklu2010}, we estimate conservatively $\mu \approx 0.2$ where the majority of the loss is due to a non-polarizing beam splitter and a near-IR objective lens.
For normally incident light and a Si refractive index of $n = 3.5$ $\trans_{Si} \approx 0.7$.
Figure~\ref{fig:Pnep} describes these sources of loss and the implied required source power to yield a low detector noise of $\sigma = 0.01$.
Our estimate for $\beta$ was taken from Figure 1 in \cite{Paniccia1998} is shown in Figure~\ref{fig:Pnep}a and corresponds to a heavily doped Intel substrate.
The required source powers assuming the same extended InGaAs detector as in Figure~\ref{fig:Pnep} and three different thickness substrates are shown in panel d. Milliwatt level source powers are clearly sufficient.

\begin{figure}[htb]
  \centering
  \includegraphics[width=0.75\textwidth]{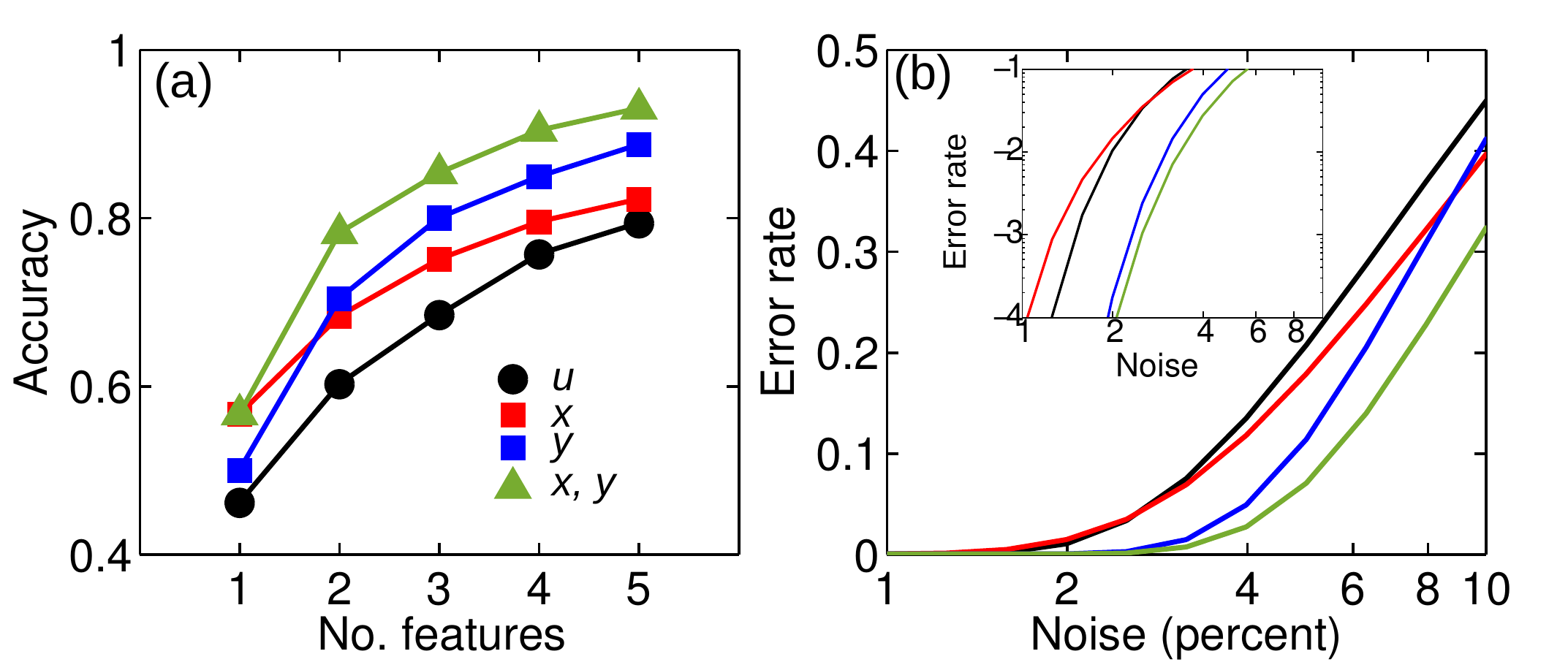}
  \caption{Influence of Gaussian noise on gate classification accuracy. 
  (a) Classification accuracy as a function of number of features used for a standard deviation of $\sigma = 0.05$. 
  (b) Error rate (1 - Accuracy) as a function of noise ($\sigma$) in percent. 
  The inset shows the low noise rates on a log scale.}
  \label{fig:class-gauss}
\end{figure}

\subsection{Gaussian Noise Approximation Estimates}
\label{app:snr:gaussian-ests}

To initially estimate the error in classification as a function of detector noise, we modeled the probability distribution at each wavelength as a Gaussian with a fixed standard deviation, $\sigma$, and mean $\mu$ given by the reflectance at a given feature,

\begin{equation} \label{eq:gauss}
  p(M_i \vert c_j) = \frac{1}{\sqrt{2\pi\sigma}}e^{-(M_i - \mu_{ij})^2/(2\sigma^2)}
\end{equation}

\noindent where $\mu_{ij} = \refl(\lambda_i, q_i, c_j)$ using the spectra in Figure 3.
Since the decision rule is $\vec{M} \in R_j \rightarrow c = c_j$, the error in classification can be calculated analytically as the overlap integral,

\begin{equation} \label{eq:error}
  K_{ij} = \int_{R_j}p(\vec{M} \vert c_i)P(c_i)\,d\vec{M}
\end{equation}

\noindent which gives the probability of incorrectly classifying gate type $c_i$ as $c_j$.
The total error is the sum over all possible combinations of incorrect classifications.
In order to compute the integral in equation~\ref{eq:error} the quantity $p(\vec{M} \vert c_i)$ still needs to be calculated just as was necessary to form the classification probability $P(c_i \vert \vec{M})$. Since we again used the assumption of independence the error rate calculated in this way is an approximation of the true error rate.
We used equation~\ref{eq:error} to determine the error in classification and selected classification features using the greedy approach described for a standard deviation of $\sigma = 0.05$.
Figure~\ref{fig:class-gauss}a shows the accuracy increase as features are added for the four polarization configurations.
We then fixed our feature vector $\vec{M}$ and calculated the error rate as the noise level was varied.
The results are shown in Figure~\ref{fig:class-gauss}b and indicate that for $\sigma < 0.03$ accuracies of more than 99\% (error below $10^{-2}$) are possible.


\begin{thebibliography}{10}
\bibitem{NA_note}
We emphasize that throughout NA refers to the \emph{total} system numerical aperture, accounting for the SIL, not that of the backing objective.
\end{thebibliography}

\end{document}